\pdfoutput=1

\documentclass[11pt,twoside,a4paper,cmspaper,final,collab]{cms-tdr}
\def\svnVersion{bf972e7-D}\def\svnDate{2019/07/11}\def\cmsCernNoTag{CERN-EP-2018-264}\def\cmsCernDate{\today}\def\cmsMessage{"Published in Physical Review Letters as \href{http://dx.doi.org/10.1103/PhysRevLett.123.022001}{\doi{10.1103/PhysRevLett.123.022001}.}"}

\begin{document}\cmsNoteHeader{HIN-16-016}

\hyphenation{had-ron-i-za-tion}
\hyphenation{cal-or-i-me-ter}
\hyphenation{de-vices}
\RCS$HeadURL$
\RCS$Id$

\newlength\cmsFigWidth
\ifthenelse{\boolean{cms@external}}{\setlength\cmsFigWidth{0.85\columnwidth}}{\setlength\cmsFigWidth{0.4\textwidth}}
\ifthenelse{\boolean{cms@external}}{\providecommand{\cmsLeft}{top\xspace}}{\providecommand{\cmsLeft}{left\xspace}}
\ifthenelse{\boolean{cms@external}}{\providecommand{\cmsRight}{bottom\xspace}}{\providecommand{\cmsRight}{right\xspace}}

\newcommand{\sqrts}{\ensuremath{\sqrt{s}}\xspace}
\newcommand{\BtoD}{\ensuremath{\PB\to\PDz}\xspace}
\newcommand{\Dzerodecay}{\ensuremath{\PDz\to\PKm\Pgpp}\xspace}
\newcommand{\RAA}{\ensuremath{R_{\mathrm{AA}}}\xspace}
\newcommand{\taa}{\ensuremath{T_{\mathrm{AA}}}\xspace}
\newcommand{\absy}{\ensuremath{ \abs{y}}\xspace}
\newcommand{\abseta}{\ensuremath{ \abs{\eta}}\xspace}
\providecommand{\PbPb}{\ensuremath{\text{PbPb}}\xspace}

\cmsNoteHeader{HIN-16-016}

\title{Studies of beauty suppression via nonprompt \texorpdfstring{\PDz}{D0} mesons in \texorpdfstring{\PbPb}{PbPb} collisions at \texorpdfstring{$\sqrtsNN=5.02\TeV$}{sqrt(s[NN]) = 5.02 TeV}}

\date{\today}

\abstract{
The transverse momentum spectra of {\PDz} mesons from {\cPqb} hadron decays are measured at midrapidity ($\absy<1$) in {\Pp}{\Pp} and \PbPb collisions at a nucleon--nucleon center of mass energy of 5.02\TeV with the CMS detector at the LHC. The {\PDz} mesons from {\cPqb} hadron decays are distinguished from prompt {\PDz} mesons by their decay topologies. In \PbPb collisions, the \BtoD yield is found to be suppressed in the measured \pt range from 2 to 100\GeVc as compared to {\Pp}{\Pp} collisions. The suppression is weaker than that of prompt {\PDz} mesons and charged hadrons for \pt around 10\GeVc. While theoretical calculations incorporating partonic energy loss in the quark-gluon plasma can successfully describe the measured \BtoD suppression at higher \pt, the data show an indication of larger suppression than the model predictions in the range of $2<\pt<5\GeVc$.
}

\hypersetup{
pdfauthor={CMS Collaboration},
pdftitle={Studies of beauty suppression via measurements of nonprompt D0 mesons in PbPb collisions at sqrt(sNN)=5.02 TeV},
pdfsubject={CMS},
pdfkeywords={CMS, physics, B meson, D meson, pT spectrum, nuclear modification factor}}

\maketitle

Quantum chromodynamics (QCD) predicts the existence of a quark-gluon plasma (QGP) phase, consisting of deconfined quarks and gluons, at extremely high temperatures and/or densities~\cite{QGP1,QGP2,Karsch:2003jg}. Experiments at the BNL RHIC and the CERN LHC indicate that a strongly coupled QGP is created in relativistic heavy ion collisions at nucleon--nucleon center-of-mass energies \sqrtsNN from 200\GeV to several \TeV~\cite{Adams:2005dq,Adcox:2004mh,Back:2004je,Arsene:2004fa,Muller:2012zq}. Heavy quarks (charm and beauty) produced in heavy ion collisions are valuable probes for studying the properties of this deconfined medium. They are mostly produced in primary hard QCD scatterings at an early stage of the collision.
During their propagation through the QGP, heavy quarks lose energy via radiative and collisional interactions with the medium constituents, with the 2 processes dominating at high and low transverse momentum (\pt), respectively.
Parton energy loss can be studied using the nuclear modification factor (\RAA), which is defined as the ratio of the particle yield in nucleus--nucleus (AA) to that in proton--proton (pp) collisions, normalized by the number of binary nucleon--nucleon collisions ($N_{\mathrm{coll}}$)~\cite{Miller:2007ri}. Precise measurements of \RAA for particles containing light, charm, and beauty quarks over a wide \pt range can test the predicted flavor (parton mass) and energy dependence of the parton energy loss in the QGP~\cite{Andronic:2015wma}. This can provide both important tests of QCD at extreme densities and temperatures, and constraints on theoretical models describing the system evolution in heavy ion collisions.

Charm suppression in heavy ion collisions was reported by RHIC and LHC experiments~\cite{PhysRevLett.113.142301,Adam:2015nna,Adam:2015sza,Abelev:2014laa,Abelev:2012hxa,HIN16001}.
For beauty production, the CMS Collaboration measured \RAA for nonprompt {\PJGy} mesons (coming from decays of {\cPqb} hadrons) and for fully reconstructed {\PBpm} mesons~\cite{Khachatryan:2016ypw,Bmesonpaper,HIN16025}. A suppression by a factor of about two was observed in both channels for $\pt>6\GeVc$ at mid-rapidity.
At the same time, the \RAA of nonprompt {\PJGy} mesons in the \pt range of 6.5--30\GeVc was found to be larger than the \RAA of prompt D mesons in the 8--16\GeVc \pt region for central events, which is in line with a mass ordering of quark energy loss~\cite{Andronic:2015wma}.
An indication of less suppression of nonprompt {\PJGy} mesons is seen at forward rapidity ($1.8<\absy<2.4$), at low \pt, down to 3\GeVc.
Extending measurements of charm and beauty suppression to a broader \pt coverage should provide improved discrimination between the radiative and collisional parton energy loss mechanisms, leading to better constraints on theoretical predictions.

In this letter, we report a study of beauty production and in-medium energy loss performed by measuring nonprompt {\PDz} \pt spectra in {\Pp}{\Pp} and 0\%--100\% centrality (i.e., the degree of overlap of the two colliding nuclei) \PbPb collisions at \sqrtsNN = 5.02\TeV with the CMS detector. The measurement is done in the rapidity region $\absy<1$, in a wide \pt range from 2 to 100\GeVc.
The {\PDz} and {\PaDz} mesons, whose yields are merged in this analysis, are reconstructed via the hadronic decay channel \Dzerodecay that has a branching fraction of 3.93\%~\cite{PDG2016}. The combined branching fractions of {\PB} mesons $\to$ {\PDz}X/{\PaDz}X and the following \Dzerodecay are significantly higher than those for previous measurements via nonprompt {\PJGy} mesons and fully reconstructed {\PBpm} mesons.

The central feature of the CMS apparatus is a superconducting solenoid of 6\unit{m} internal diameter, providing a magnetic field of 3.8\unit{T}. Within the solenoid volume are a silicon pixel and strip tracker, a lead tungstate crystal electromagnetic calorimeter, and a brass and scintillator hadron calorimeter, each composed of a barrel and two endcap sections.
The silicon tracker measures charged particles within the pseudorapidity range $\abseta<2.5$.
For nonisolated particles of $1<\pt<10\GeVc$ and $\abseta<1.4$, the track resolutions are typically 1.5\% in \pt and 25--90 (45--150)\mum in the transverse (longitudinal) impact parameter \cite{TRK-11-001}.
A detailed description of the CMS experiment can be found in Ref.~\cite{bib_CMS}.

This analysis is performed using {\Pp}{\Pp} and \PbPb data collected in 2015 at \sqrtsNN= 5.02\TeV.
For {\PDz} \pt less than 20\GeVc, minimum-bias samples corresponding to about 2.67 billion {\Pp}{\Pp} (294 million \PbPb) collisions are used.
For {\PDz} \pt above 20\GeVc, we use samples from dedicated {\PDz} high-level trigger (HLT) algorithms~\cite{HIN16001}, corresponding to integrated luminosities of 27.4\pbinv~\cite{CMS-PAS-LUM-16-001} and 530\mubinv for {\Pp}{\Pp} and \PbPb collisions, respectively.
The same event selection as in Refs.~\cite{Khachatryan:2016odn,Khachatryan:2010xs,HIN16001} is used to reject instrumental background processes (beam--gas collisions, beam scraping events and ultra-peripheral non-hadronic collisions).

Monte Carlo (MC) simulated events are used to evaluate detector acceptance, reconstruction and selection efficiency for {\PDz}, and to obtain geometrical distributions for prompt and nonprompt {\PDz} meson decay vertices relative to the primary vertex (PV, the reconstructed collision point).
The MC samples are produced by generating {\Pp}{\Pp} collisions containing a {\PDz} meson with \PYTHIA 8.122~\cite{Sjostrand:2014zea} tune CUETP81M1~\cite{Khachatryan:2015pea}. The decay kinematics of the heavy flavor hadrons are simulated with \EVTGEN 1.3.0~\cite{evtgen}. Each {\Pp}{\Pp} event is then overlaid with a \PbPb collision event generated with \HYDJET 1.8~\cite{Lokhtin:2005px}. The centrality distribution in real data is approximated by weighting the \HYDJET event sample by the number of inelastic nucleon-nucleon collisions. The generated {\PB} meson \pt distributions are also weighted such that they reproduce the measured nonprompt {\PDz} spectra in this analysis. The detector response is simulated with \GEANTfour~\cite{geant4}.

The {\PDz} candidates are reconstructed by combining pairs of oppositely charged tracks.
Each track is required to pass a high purity selection based on a multi-variate analysis of track quality variables~\cite{CMS-PAS-TRK-10-001}. Tracks are required to have $\abseta<1.5$ and \pt larger than 1 \GeVc for the {\Pp}{\Pp} and \PbPb minimum-bias data, and 2 and 8.5\GeVc for {\Pp}{\Pp} and \PbPb \PDz-triggered samples, respectively.
For each pair of selected tracks, two {\PDz} candidates are created by assuming that one of the particles has the pion mass and the other has the kaon mass, and vice-versa. The {\PDz} candidates are required to have $\absy<1$, where the track resolution is better.
In order to reduce the combinatorial background and prompt {\PDz} contribution, the {\PDz} candidates are selected based on several geometrical criteria:
a minimum probability that the two tracks come from a common decay vertex,
a minimum distance between the decay vertex and the PV divided by its uncertainty,
and minimum distances of closest approach (DCA) to the PV for the pion and kaon tracks divided by their uncertainties.
The selection is optimized using simulated signal samples complemented by background events from mass sidebands in the data. Dedicated optimizations are performed for different \pt ranges and for {\Pp}{\Pp} and \PbPb collisions, in order to maximize the statistical significance of the \BtoD (\ie, {\PDz} mesons from {\cPqb} hadron decays) yield.

\begin{figure}[h]
  \begin{center}
    \includegraphics[width=0.48\textwidth]{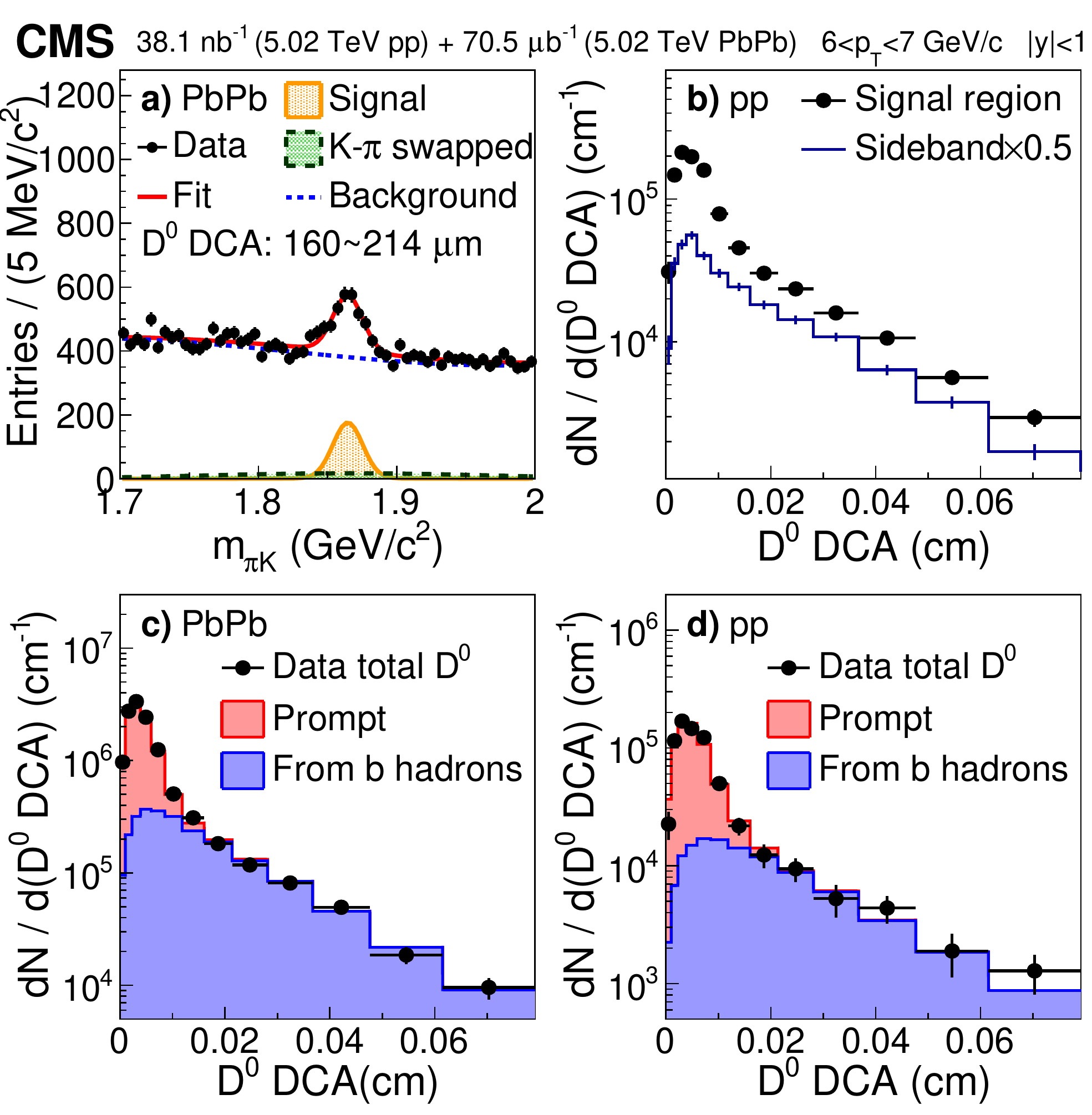}
    \caption{a) Example of a three-component invariant mass fit of a {\PDz} DCA bin for \pt of 6--7\GeVc in \PbPb collisions;
b) DCA distributions for {\PDz} candidates in the signal invariant mass region and in the sidebands (scaled by the mass range ratio of 0.5) for {\PDz} \pt of 6--7\GeVc in {\Pp}{\Pp} collisions;
c) Signal DCA distribution obtained with the invariant mass fit for each DCA bin, and a prompt+nonprompt two-component fit to it, for {\PDz} \pt of 6--7\GeVc in \PbPb collisions;
d) Signal DCA distribution obtained with the sideband subtraction, and a prompt+nonprompt two-component fit to it, for {\PDz} \pt of 6--7\GeVc in {\Pp}{\Pp} collisions.}
    \label{fig:InvMSideBandFits}
  \end{center}
\end{figure}

The \BtoD decays are distinguished from prompt {\PDz} mesons by fitting the distribution of DCA between the {\PDz} path and the PV.
The signal {\PDz} DCA distribution, including both the prompt and nonprompt components, is extracted by two methods. For \pt bins in which there is abundant background ({\PDz} $\pt<20\GeVc$ for \PbPb), the {\PDz} meson yield in each {\PDz} DCA bin is obtained from an invariant mass fit with three components: a double-Gaussian function describing the signal, a broad Gaussian function describing \PK--\PGp swapped pairs, and a third-order polynomial component for the combinatorial background. Figure~\ref{fig:InvMSideBandFits}a shows an example of a three-component invariant mass fit for a selected {\PDz} DCA and \pt bin.
For the {\Pp}{\Pp} data and for {\PDz} candidates with $\pt>20\GeVc$ from \PbPb events, for which the background is low, a sideband subtraction method is used to obtain the signal {\PDz} DCA distribution.
Figure~\ref{fig:InvMSideBandFits}b shows the DCA distributions for {\PDz} candidates in the signal invariant mass region ($\abs{m_{\text{rec}}-m_{\PDz}}<0.025\GeVcc$) and for candidates in the sidebands ($0.05<\abs{m_{\text{rec}}-m_{\PDz}}<0.1\GeVcc$). The latter is scaled by the mass range ratio of 0.5, in order to estimate the background yield in the narrower signal region.
Here $m_{\mathrm{rec}}$ is the reconstructed \PK--\PGp{} invariant mass and $m_{\PDz}$ is the nominal mass of the {\PDz} meson, 1.8648\GeVcc~\cite{PDG2016}.
The signal {\PDz} DCA distribution is calculated as the difference of the {\PDz} DCA distributions in the signal region and the sidebands.

In order to obtain the \BtoD yield, a two-component fit to the signal {\PDz} DCA distribution is carried out using prompt and nonprompt {\PDz} DCA templates obtained from MC simulations, as shown in Fig.~\ref{fig:InvMSideBandFits}c and~\ref{fig:InvMSideBandFits}d, for \PbPb and pp, respectively.
The prompt {\PDz} mesons have a narrow DCA distribution near zero, with the width purely resulting from the detector resolution, while the nonprompt {\PDz} DCA distribution is much wider because of the kink between the {\cPqb} hadron and {\PDz} meson directions.
This two-component fit is sensitive to the modeling of the {\PDz} DCA distributions in the simulation.
To assess systematic effects on the two-component fit arising from potential differences between the resolution in data and simulation, the widths of the simulated DCA distributions are varied by a floating scale factor.
The best simulated DCA width scale factor to match the data is determined by minimizing the $\chi^2$ of the two-component fit. It is found to be in the range of $1.0\pm0.1$ for all \pt bins, indicating a good data-to-simulation consistency.

The \BtoD differential cross section with $\absy<1$ in {\Pp}{\Pp} collisions is calculated with the following equation:
\begin{equation}
  \left.\frac{\rd\sigma^{\BtoD}_{{\Pp}{\Pp}}}{\rd\pt}\right|_{\abs{y}<1} =
  \frac{1}{2\mathcal{L}\Delta\pt\mathcal{B}}\left.\frac{N^{\BtoD+\PaDz}_{{\Pp}{\Pp}}}{\alpha\epsilon}\right|_{\abs{y}<1}.
  \label{eq:crosssectionPP}
\end{equation}
Here $N^{{\BtoD+\PaDz}}_{{\Pp}{\Pp}}$ are the nonprompt {\PDz} and {\PaDz} meson yields extracted in each \pt interval; $\mathcal{L}$ is the integrated luminosity for the corresponding trigger; $\Delta \pt$ is the width of the \pt interval; $\mathcal{B}$ is the decay branching fraction; and $\alpha\epsilon$ represents the product of acceptance and efficiency.
The factor 1/2 accounts for the fact that the yields were measured for {\PDz} plus {\PaDz}, but the cross section is for either {\PDz} or {\PaDz} production.

The \BtoD yield with $\absy<1$ in \PbPb collisions is calculated similarly, and normalized by the nuclear overlap function $\taa = N_{\text{coll}} / \sigma_{\mathrm{NN}}^{\text{inelastic}} = 5.61\mbinv$~\cite{Khachatryan:2016odn} calculated with the Glauber model~\cite{Miller:2007ri}, to facilitate the comparison with the {\Pp}{\Pp} spectrum, as:
\begin{linenomath}
\ifthenelse{\boolean{cms@external}}
{ 
\begin{multline}
  \frac{1}{\taa}\left.\frac{\rd N^{\BtoD}_{\PbPb}}{\rd\pt}\right|_{\abs{y}<1} \\
  = \frac{1}{\taa}\frac{1}{2N_{\text{events}}\Delta\pt\mathcal{B}}\left.\frac{N^{\BtoD+\PaDz}_{\PbPb}}{\alpha\epsilon}\right|_{\abs{y}<1},
  \label{eq:invYieldPbPb}
\end{multline}
} 
{ 
\begin{equation}
  \frac{1}{\taa}\left.\frac{\rd N^{\BtoD}_{\PbPb}}{\rd\pt}\right|_{\abs{y}<1}
  = \frac{1}{\taa}\frac{1}{2N_{\text{events}}\Delta\pt\mathcal{B}}\left.\frac{N^{\BtoD+\PaDz}_{\PbPb}}{\alpha\epsilon}\right|_{\abs{y}<1},
  \label{eq:invYieldPbPb}
\end{equation}
} 
\end{linenomath}
where the number of sampled inelastic collision events $N_{\text{events}}$ replaces the integrated luminosity $\mathcal{L}$.

The nuclear modification factor is defined as
\begin{equation}
  R_{\mathrm{AA}} = \frac{1}{\taa} \frac{\rd N^{\BtoD}_{\PbPb}}{\rd\pt} / \frac{\rd\sigma^{\BtoD}_{{\Pp}{\Pp}}}{\rd\pt}.
  \label{eq:RAA}
\end{equation}
The global systematic uncertainty (common to all points) of the \BtoD \pt spectrum in {\Pp}{\Pp} collisions (2.5\%) is the sum in quadrature of the uncertainties in the integrated luminosity (2.3\%~\cite{CMS-PAS-LUM-16-001}) and in the \Dzerodecay branching fraction (1\%~\cite{PDG2016}).
The global uncertainty in the \PbPb measurement ($+$4.1\%, $-$3.6\%) includes the uncertainties in the number of sampled \PbPb inelastic collision events (2\%), in the branching fraction (1\%), and in $\taa$ ($+$2.8\%, $-$3.4\%~\cite{Khachatryan:2016odn}).
In the calculation of \RAA, the uncertainty in the branching fraction cancels out.
The other uncertainties are summed in quadrature, amounting to a total global systematic uncertainty in the \RAA of $+$4.6\%, $-$4.1\%.

The following systematic uncertainties are evaluated separately in different \pt ranges.
The systematic uncertainty due to the signal extraction from the invariant mass fit (3.2--5.3\%) is evaluated by varying the function used to fit the background, and by comparing the default double-Gaussian signal yield with that obtained with a different method, in which the integral of a third-order polynomial function describing the background and the \PK--\PGp swapped pairs in the signal invariant mass region is subtracted from the number of candidate counts. 
The uncertainty due to the signal extraction with the sideband subtraction method (1.4--8.6\%) is obtained by comparing the {\PDz} meson yield from the sideband method with the yield from the invariant mass fit, both obtained within the {\PDz} DCA range where the nonprompt {\PDz} component dominates.
The systematic uncertainty associated with the separation of prompt {\PDz} mesons and {\PDz} mesons from {\cPqb} hadron decays (4.2--30.4\%) comes from two sources. The first part, which is due to the data--simulation difference in the {\PDz} DCA shapes, is estimated by comparing the default \BtoD yields (from the two-component fit using MC DCA templates with varied widths to match the data) with that obtained using the original MC DCA templates without the width variation. The second part, which is due to statistical uncertainty in the simulated samples, is obtained by smearing simulated {\PDz} DCA distributions according to the statistical uncertainties in each individual bin, and repeating the two-component fit 1000 times.
The systematic uncertainty in the tracking efficiency is 4\% for a single track~\cite{CMS-PAS-TRK-10-002}, and 8\% for a pair of tracks. For \RAA, the systematic uncertainty in the tracking efficiency ratio between \PbPb and {\Pp}{\Pp} data is 6\% for a track~\cite{Khachatryan:2016odn}, and 12\% for a pair of tracks.
The systematic uncertainty in the selection efficiency due to the geometrical criteria (6.9--11.6\%) is evaluated by varying the selection variables.
The systematic uncertainty in the {\PDz} HLT trigger efficiency (2.0--7.9\%) is from the statistical precision of the number of {\PDz} meson candidates in the events common to the {\PDz} triggered and minimum-bias triggered samples.
The systematic uncertainty in the acceptance and efficiency due to the simulated {\PB} meson \pt distribution (0.0--3.6\%) is estimated by changing the default {\PB} meson \pt shapes (that reproduce the measured nonprompt {\PDz} spectra) to the fixed-order next-to-leading logarithm (FONLL)~\cite{FONLL} pQCD calculated (pp) and FONLL+TAMU model~\cite{He:2011qa, He:2014cla} predicted (PbPb) {\PB} meson \pt shapes.
The systematic uncertainty in the acceptance and efficiency due to the simulated {\PB} meson centrality distribution (0.4--2.3\%) is estimated by assuming the {\PB} meson yield to be proportional to the number of participating nucleons instead of the number of inelastic nucleon-nucleon collisions.
The total systematic uncertainty in each \pt interval is computed as the sum in quadrature of the individual uncertainties listed above.

\begin{figure}[h]
\begin{center}
\includegraphics[width= 0.48\textwidth]{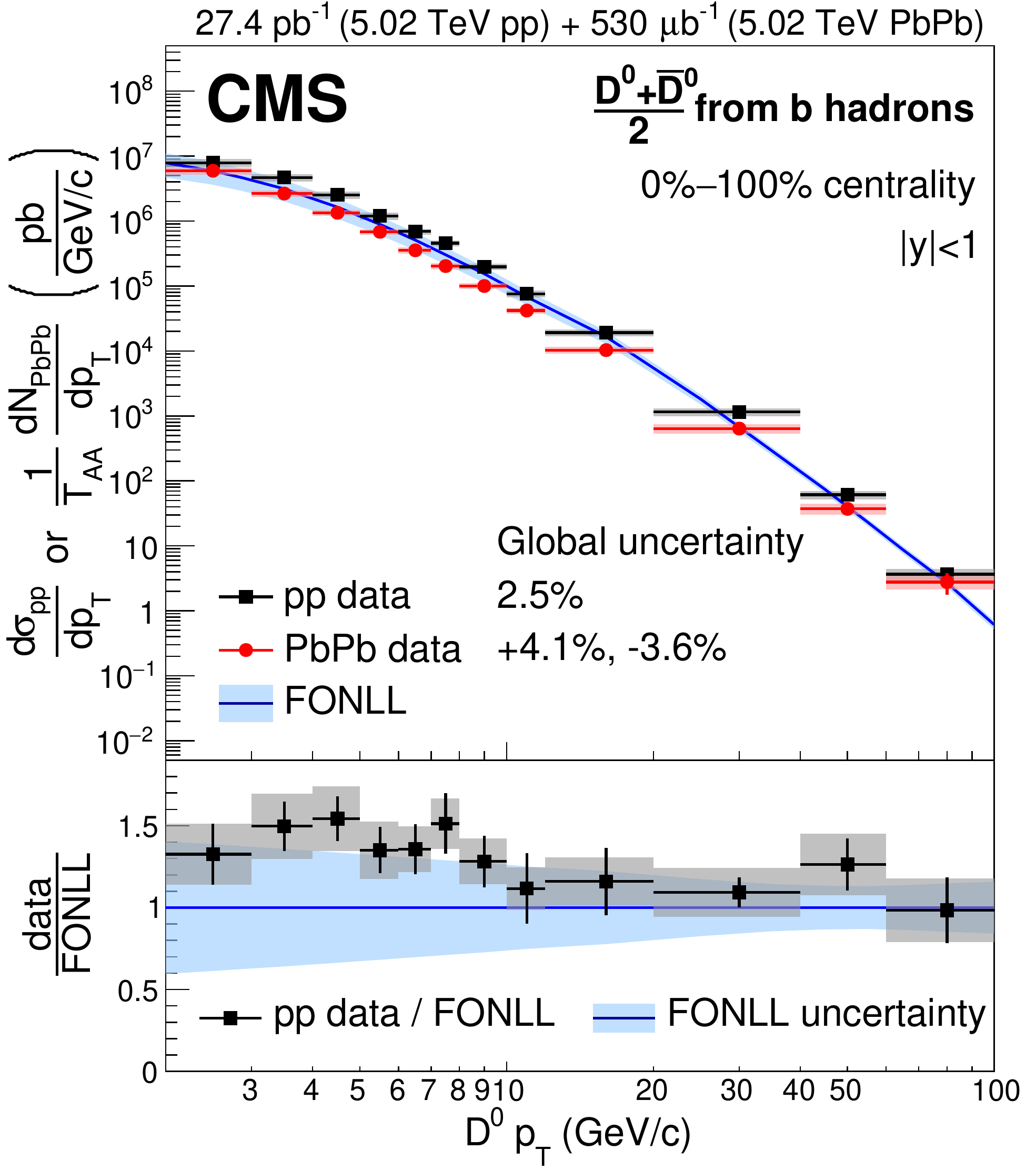}
\caption{Upper panel: \BtoD \pt-differential cross section in {\Pp}{\Pp} collisions and invariant yield in \PbPb collisions normalized with \taa, at \sqrtsNN= 5.02\TeV. The vertical bands around the data points represent the bin-by-bin systematic uncertainties. Uncertainties are smaller than the symbols in most cases. The cross section in {\Pp}{\Pp} collisions is compared to FONLL calculations~\cite{FONLL}. Lower panel: The $\text{data}/\mathrm{FONLL}$ ratio for the \BtoD \pt spectra in {\Pp}{\Pp} collisions.}
\label{fig:ptSpectra}
\end{center}
\end{figure}

In Fig.~\ref{fig:ptSpectra}, the \BtoD \pt-differential cross section in {\Pp}{\Pp} collisions and the invariant yield in \PbPb collisions normalized with \taa are presented.
The plot also shows the nonprompt {\PDz} \pt spectra found by decaying a {\PB} meson \pt spectrum calculated using FONLL~\cite{FONLL} pQCD.
The ratio of the measured {\Pp}{\Pp} spectrum over the FONLL prediction is shown in the bottom panel. The measurement in {\Pp}{\Pp} collisions lies close to the upper limit of the FONLL predicted range.

\begin{figure}[h]
\begin{center}
\includegraphics[width= 0.48\textwidth]{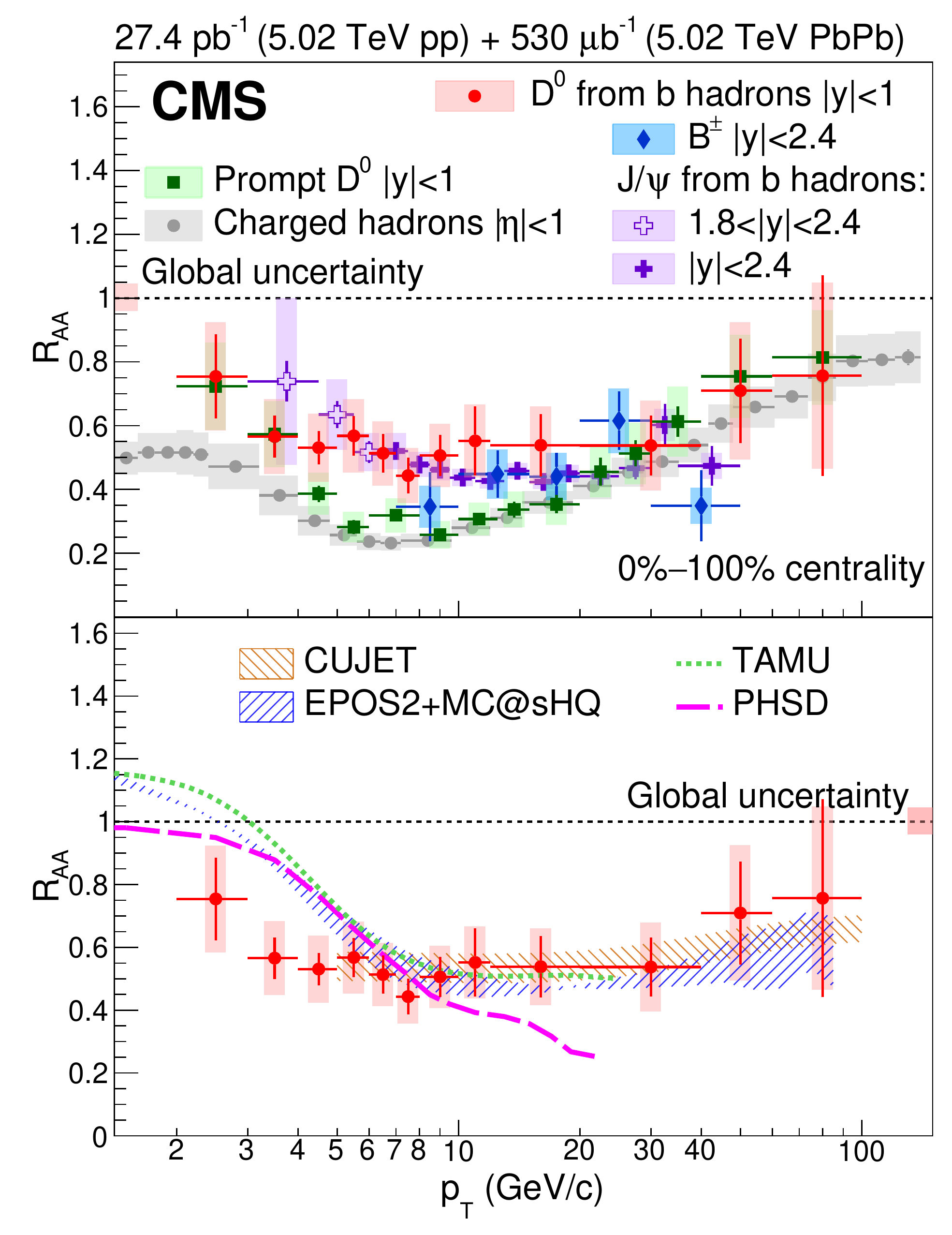}
\caption{The \BtoD nuclear modification factor \RAA for \PbPb collisions at \sqrtsNN= 5.02\TeV (red circles) compared to other particles~\cite{Bmesonpaper,HIN16025,HIN16001,Khachatryan:2016odn} (upper panel), and to various theoretical predictions~\cite{Xu:2015bbz, Xu:2014ica,Xu:2014tda,Gossiaux:2014jga,He:2011qa,He:2014cla,Song:2015sfa,Song:2015ykw} (lower panel). The vertical bands around the data points and at unity represent the bin-by-bin and global systematic uncertainties, respectively.}
\label{fig:RAA}
\end{center}
\end{figure}

Figure~\ref{fig:RAA} shows the \BtoD nuclear modification factor \RAA.
It can be seen that the \BtoD \RAA is below unity in the measured \pt range from 2 to 100\GeVc.
In the upper panel, the \BtoD \RAA is compared with the \RAA of {\PB} mesons~\cite{Bmesonpaper}, nonprompt {\PJGy} mesons from {\cPqb} hadron decays~\cite{HIN16025}, prompt {\PDz} mesons~\cite{HIN16001}, and charged hadrons~\cite{Khachatryan:2016odn}.
The \BtoD \RAA is close to the {\PB} meson and nonprompt {\PJGy} meson results, and extends the reach of {\cPqb} quark related \RAA studies to a larger \pt coverage at midrapidity.
The \BtoD yield is less suppressed than prompt {\PDz} mesons and charged hadrons with \pt around 10\GeVc.
This may reflect a dependence of the suppression effects on the quark mass~\cite{Andronic:2015wma}, although a direct comparison requires a full modeling of the quark initial spectrum and hadronization, as well as of the decay kinematics.

In the lower panel of Fig.~\ref{fig:RAA}, the measured \BtoD \RAA is compared with various theoretical predictions.
The \textsc{cujet} and \textsc{epos2+MC@sHQ} models are perturbative QCD-based calculations that include both collisional and radiative energy loss~\cite{Xu:2015bbz, Xu:2014ica,Xu:2014tda,Gossiaux:2014jga}.
The TAMU model is a transport model based on a Langevin equation that includes collisional energy loss and heavy quark diffusion in the medium~\cite{He:2011qa, He:2014cla}.
The PHSD model is a microscopic off-shell transport model based on a Boltzmann approach that includes collisional energy loss only~\cite{Song:2015sfa, Song:2015ykw}.
At higher \pt, the \textsc{cujet}, \textsc{epos2+MC@sHQ} and TAMU models all match the data well.
However, at \pt below 5\GeVc, our measurements show a hint of stronger suppression than predicted by all available models in this \pt range. This could indicate a stronger energy loss of {\cPqb} quarks in QGP than predicted at low \pt, where collisional parton energy loss begins to dominate. It could also be due to other effects. For example, the fraction of {\cPqb} baryons out of all {\cPqb} hadrons may be enhanced at low \pt in \PbPb collisions, because {\cPqb} quarks can hadronize by coalescing with light quarks in the medium~\cite{Das:1977cp,Fries:2008hs,Abelev:2006jr,Oh:2009zj}.
Given the much lower decay fractions of {\cPqb} baryons $\to\PDz$ with respect to the $\PBpm\to\PDz$ and $\PBz\to\PDz$ cases, fewer {\cPqb} hadrons are seen in this analysis than expected by the models.
This baryon enhancement effect is not accounted for by the models considered.

In summary, this letter presents the transverse momentum spectra of {\PDz} mesons from {\cPqb} hadron decays measured in {\Pp}{\Pp} and \PbPb collisions at a center-of-mass energy \sqrtsNN = 5.02\TeV per nucleon pair with the CMS detector at the LHC. The {\PDz} mesons from {\cPqb} hadron decays are distinguished from the prompt {\PDz} mesons by the distance of closest approach of the {\PDz} path relative to the primary vertex.
The measured spectrum in {\Pp}{\Pp} collisions is close to the upper limit of a Fixed-Order Next-to-Leading Logarithm perturbative quantum chromodynamics calculation.
In \PbPb collisions, the \BtoD yield is suppressed in the measured transverse momentum (\pt) range from 2 to 100\GeVc. The \BtoD nuclear modification factor \RAA is higher than for prompt {\PDz} mesons and charged hadrons around 10\GeVc, which is in line with a quark mass ordering of suppression. Compared to theoretical predictions, the measured \RAA is consistent with some models at higher \pt, but shows a hint of stronger suppression than all of the available models at low \pt. This could indicate a stronger energy loss of {\cPqb} quarks in the quark-gluon plasma than predicted at low \pt, or could reflect an enhanced {\cPqb} baryon production due to quark coalescence in \PbPb collisions.

\begin{acknowledgments}
We congratulate our colleagues in the CERN accelerator departments for the excellent performance of the LHC and thank the technical and administrative staffs at CERN and at other CMS institutes for their contributions to the success of the CMS effort. In addition, we gratefully acknowledge the computing centers and personnel of the Worldwide LHC Computing Grid for delivering so effectively the computing infrastructure essential to our analyses. Finally, we acknowledge the enduring support for the construction and operation of the LHC and the CMS detector provided by the following funding agencies: BMBWF and FWF (Austria); FNRS and FWO (Belgium); CNPq, CAPES, FAPERJ, FAPERGS, and FAPESP (Brazil); MES (Bulgaria); CERN; CAS, MoST, and NSFC (China); COLCIENCIAS (Colombia); MSES and CSF (Croatia); RPF (Cyprus); SENESCYT (Ecuador); MoER, ERC IUT, and ERDF (Estonia); Academy of Finland, MEC, and HIP (Finland); CEA and CNRS/IN2P3 (France); BMBF, DFG, and HGF (Germany); GSRT (Greece); NKFIA (Hungary); DAE and DST (India); IPM (Iran); SFI (Ireland); INFN (Italy); MSIP and NRF (Republic of Korea); MES (Latvia); LAS (Lithuania); MOE and UM (Malaysia); BUAP, CINVESTAV, CONACYT, LNS, SEP, and UASLP-FAI (Mexico); MOS (Montenegro); MBIE (New Zealand); PAEC (Pakistan); MSHE and NSC (Poland); FCT (Portugal); JINR (Dubna); MON, RosAtom, RAS, RFBR, and NRC KI (Russia); MESTD (Serbia); SEIDI, CPAN, PCTI, and FEDER (Spain); MOSTR (Sri Lanka); Swiss Funding Agencies (Switzerland); MST (Taipei); ThEPCenter, IPST, STAR, and NSTDA (Thailand); TUBITAK and TAEK (Turkey); NASU and SFFR (Ukraine); STFC (United Kingdom); DOE and NSF (USA).
\end{acknowledgments}

\bibliography{auto_generated}

\cleardoublepage \appendix\section{The CMS Collaboration \label{app:collab}}\begin{sloppypar}\hyphenpenalty=5000\widowpenalty=500\clubpenalty=5000\vskip\cmsinstskip
\textbf{Yerevan Physics Institute, Yerevan, Armenia}\\*[0pt]
A.M.~Sirunyan, A.~Tumasyan
\vskip\cmsinstskip
\textbf{Institut f\"{u}r Hochenergiephysik, Wien, Austria}\\*[0pt]
W.~Adam, F.~Ambrogi, E.~Asilar, T.~Bergauer, J.~Brandstetter, M.~Dragicevic, J.~Er\"{o}, A.~Escalante~Del~Valle, M.~Flechl, R.~Fr\"{u}hwirth\cmsAuthorMark{1}, V.M.~Ghete, J.~Hrubec, M.~Jeitler\cmsAuthorMark{1}, N.~Krammer, I.~Kr\"{a}tschmer, D.~Liko, T.~Madlener, I.~Mikulec, N.~Rad, H.~Rohringer, J.~Schieck\cmsAuthorMark{1}, R.~Sch\"{o}fbeck, M.~Spanring, D.~Spitzbart, A.~Taurok, W.~Waltenberger, J.~Wittmann, C.-E.~Wulz\cmsAuthorMark{1}, M.~Zarucki
% ...
\vskip\cmsinstskip
% ...
1:  Also at Vienna University of Technology, Vienna, Austria\\
% ...
\end{sloppypar}
\end{document}